\documentclass[pre,twocolumn,superscriptaddress,showpacs]{revtex4-1}

\usepackage{graphicx}
\usepackage{dcolumn}
\usepackage{bm}
\usepackage{accents}
\usepackage{color}
\newcommand{\beq}{\begin{equation}}
\newcommand{\eeq}{\end{equation}}
\newcommand{\beqa}{\begin{eqnarray}}
\newcommand{\eeqa}{\end{eqnarray}}
\newcommand{\Tr}{\text{Tr}}

\newcommand{\av}[1]{\left\langle #1 \right\rangle}

\begin{document}


\title{Path integral approach to quantum thermodynamics}


\author{Ken Funo}
\affiliation{School of Physics, Peking University, Beijing 100871, China}
\author{H. T. Quan}
\email{htquan@pku.edu.cn}
\affiliation{School of Physics, Peking University, Beijing 100871, China}
\affiliation{Collaborative Innovation Center of Quantum Matter, Beijing 100871, China}
\date{\today}

\begin{abstract}
Work belongs to the most basic notions in thermodynamics but it is not well understood in quantum systems, especially in open quantum systems. By introducing a novel concept of work functional along individual Feynman path, we invent a new approach to study thermodynamics in the quantum regime. 
Using the work functional, we derive a path-integral expression for the work statistics.  
By performing the $\hbar$ expansion, we analytically prove the quantum-classical correspondence of the work statistics. In addition, we obtain the quantum correction to the classical fluctuating work. We can also apply this approach to an open quantum system in the strong coupling regime described by the quantum Brownian motion model. This approach provides an effective way to calculate the work in open quantum systems by utilizing various path integral techniques. As an example, we calculate the work statistics for a dragged harmonic oscillator in both isolated and open quantum systems.
\end{abstract}


\maketitle

Path integral formalism of quantum mechanics and quantum field theory~\cite{FeynmanHibbs} has greatly influenced the theoretical developments of physics. It has an elegant structure for treating gauge-invariant theories.  The semi-classical limit of quantum mechanics and instantons~\cite{Rajaraman} (the tunneling effect) can be intuitively understood in this formalism. Quantum anomalies (e.g., chiral anomaly) naturally arise from the path-integral measure~\cite{Fujikawa}. Path integral allows us to understand continuous quantum phase transitions in $d$ dimensional system from a mapped $d+1$ dimensional classical system~\cite{Sondhi}. A path integral description of open quantum systems~\cite{Feynman} has been used to study the dissipative dynamics of the quantum systems, known as the Caldeira-Leggett model of the quantum Brownian motion~\cite{Caldeira83}.

Quantum thermodynamics~\cite{Hanggiaspects,Esposito,Anders,Pekola,Strasberg,fluctuation1} is an emergent field studying the nonequilibrium statistical mechanics of the quantum dissipative systems~\cite{CaldeiraBook,Sun94,Hanggi05C}. Topics in this field include the role of coherence and entanglement in the heat transfer in quantum devices~\cite{Ueda,Saito,Kato} and in the quantum heat engines~\cite{Dong,coherence} and refrigerators~\cite{Pekola16}. Quite recently, experimental studies have been put forward, such as the experimental verification of the exact nonequilibrium relations~\cite{quantumjarexp} and the implementation of the quantum Maxwell demon~\cite{Benjamin,Masuyama}. Connections to quantum information theory have been explored extensively in the studies of Maxwell demon~\cite{Parrondo} and resource theories~\cite{Horodecki}. Previous efforts of constructing a framework of quantum thermodynamics were mainly based on operator formalisms. For example, in Refs.~\cite{Campisi09,fluctuation1}, the composite system is treated as an isolated system, but the definition of fluctuating work via two-point energy measurements over the composite system is thought to be ad hoc. 
In Refs.~\cite{Horowitz1,Hekking,Liu1,Suomela}, a framework based on the quantum jump method, which was borrowed from quantum optics, is established. However, this framework is restricted to very limited cases: the weak-coupling, Markovian and rotating-wave approximation (RWA) regime. 
Hence, how to understand quantum work~\cite{Hanggiaspects} (including relations to its classical counterpart) and calculate its distributions in generic open quantum systems become the most challenging problems in this field.

\begin{figure*}[t]
\begin{center}
\includegraphics[width=.8\textwidth]{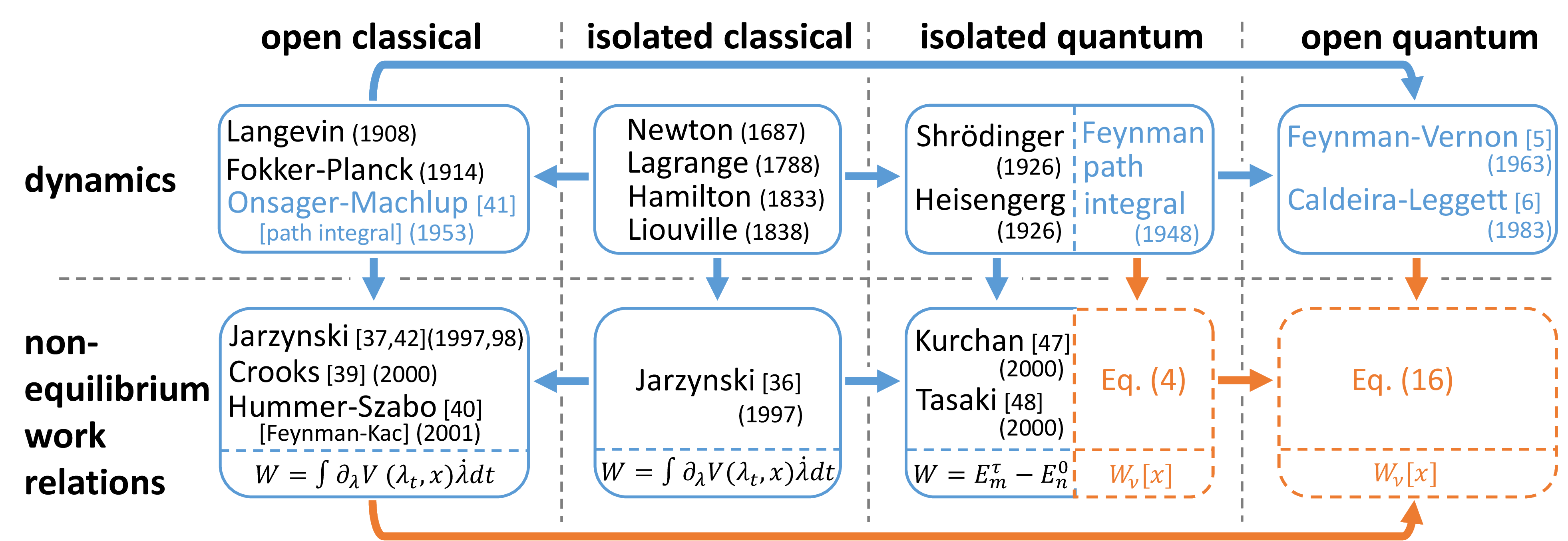}
\caption{ Summary of historical developments of theories of dynamics and nonequilibrium work relations. The arrow between every two boxes indicates the order in which the theories were developed. The fluctuating work $W$ along individual trajectories is defined differently in different contexts (see the bottom line of every column). We introduce the quantum work functional $W_{\nu}[x]$~(\ref{WFunction}) along individual Feynman paths and invent path integral approach to study quantum thermodynamics (orange dashed box). Note that Feynman's path~\cite{FeynmanHibbs} is an analogue of the Onsager-Machlup's stochastic trajectory~\cite{Onsarger} (blue color). Similarly, our definition of the work functional along individual Feynman path is an analogue of Sekimoto and Jarzynski's definition of the work functional as an integral of the supplied power~\cite{Hanggiaspects}.
}
\label{fig:classification}
\end{center}
\end{figure*}

Classical stochastic thermodynamics~\cite{Sekimoto,Sekimoto98,Seifert,Stochasticbook}, on the other hand, is a framework established in the past two decades, which extends the principles of thermodynamics from ensemble level to individual trajectory level. For example, work, heat and entropy production are identified as trajectory functionals. The first law is reformulated on the trajectory level, and the second law is refined from inequalities to equalities, known as fluctuation theorems (FT)~\cite{Jarzynski1,Jarzynski2,Crooks,Crooks00,Hummer}. The Onsarger-Machlup  ``path integral" approach~\cite{Onsarger} in formulating the FT~\cite{cpath0,cpath1,cpath2,cpath3} in classical stochastic thermodynamics is an analogue~\cite{Kac} of Feynman's path integral formalism in quantum mechanics~\cite{FeynmanHibbs}. Thus, when extending the classical stochastic thermodynamics to quantum regime, a natural idea is to do it based on path integral methods. Nevertheless, no attempt to reformulate quantum FT through Feynman's path integral formalism has succeeded so far (see Fig.~\ref{fig:classification} for historical developments in relevant fields).

In this Letter, we introduce a quantum work functional along individual Feynman path in quantum systems, and study quantum work statistics. For isolated quantum systems we reformulate the FT (the Jarzynski equality)~\cite{Tasaki,Kurchan} through path integral approach. For the open quantum system, 
we study work statistics and FT based on path integral methods~\cite{Caldeira87,Weiss,Grabert} (see Fig.~\ref{fig:classification}). In particular, we can study the non-Markovian, non-RWA, and strong coupling regime without making any approximations~\cite{Tanimura}. This is intriguing since stochastic thermodynamics~\cite{Jarzynski04,Jarzynski17,Seifert16,Talkner16} and quantum thermodynamics~\cite{Hu,Aurell,Aurell17,Carrega15,Carrega16} with strong coupling have attracted much attention recently. We utilize the semi-classical approximation technique of the path integral and show the quantum-classical correspondence of the work statistics. Furthermore, quantum corrections to the classical work functional is obtained, bringing new insights into our understandings about quantum effects in thermodynamics.

\noindent{\it Two-point measurement scheme.---} We first consider an isolated system with the system Hamiltonian given by $H_{\rm{S}}(\lambda_{t})=\hat{p}^{2}/(2M)+\hat{V}(\lambda_{t},\hat{x})$, where $M$ is the mass and $\hat{V}(\lambda_{t},\hat{x})$ is an arbitrary potential, whose time-dependence is specified by $\lambda_{t}$. This external control of the potential drives the system out of equilibrium and injects work into the system. The  fluctuating work in an isolated system is defined via the so-called two-point measurement scheme~\cite{Tasaki,Kurchan}. By measuring the energy of the system twice ($E_{n}^{0}$ and $E_{m}^{\tau}$) at $t=0$ and $t=\tau$, we define the quantum fluctuating work as the difference in the measured energies: $W_{m,n}:=E_{m}^{\tau}-E_{n}^{0}$. The joint probability about observing such measured energies is given by $p(n,m):=p_{n}|\langle m(\tau)|U_{\rm{S}}|n(0)\rangle|^{2}$, where $p_{n}:=\langle n(0)|\rho_{\rm{S}}(0)|n(0)\rangle$, $\rho_{\rm{S}}(0):=e^{-\beta H_{\rm{S}}(\lambda_{0})}/Z_{\rm{S}}(\lambda_{0})$ is the initial canonical density matrix of the system at the inverse temperature $\beta$, $|n(t)\rangle$ is the $n$-th instantaneous energy eigenstate of the system at time $t$, and $U_{\rm{S}}:=\hat{\text{T}}[\exp[(-i/\hbar)\int^{\tau}_{0}dt H_{\rm{S}}(\lambda_{t})]]$ is the unitary operator describing the time evolution of the system. The work probability distribution is given by $P(W):=\sum_{m,n}\delta(W-W_{m,n})p(m,n)$. Taking the Fourier transformation of the work probability distribution, we define the characteristic function of work~\cite{characteristic1} by $\chi_{W}(\nu):=\int dW P(W)e^{i\nu W}$. This can be expressed as
\beq
\chi_{W}(\nu)=\text{Tr}[U_{\rm{S}}e^{-i\nu H_{\rm{S}}(\lambda_{0})}\rho_{\rm{S}}(0)U_{\rm{S}}^{\dagger}e^{i\nu H_{\rm{S}}(\lambda_{\tau})}]. \label{isoCF}
\eeq

\begin{figure}[t]
\begin{center}
\includegraphics[width=.45\textwidth]{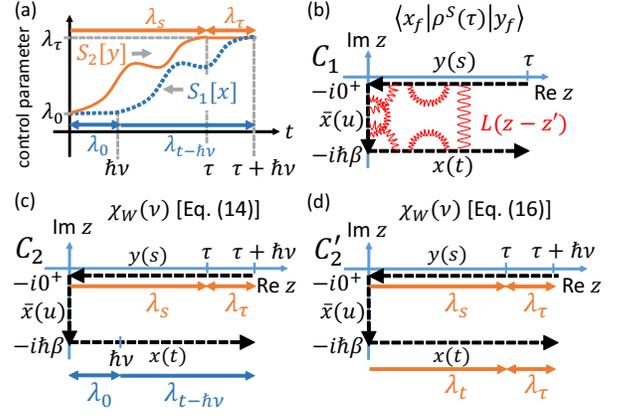}
\caption{ Contours used in the path integral and the time-dependence of the actions. (a) Time-dependence of the external control used in $S_{1}^{\nu}[x]$ (dotted blue curve) and $S_{2}^{\nu}[y]$ (solid orange curve). (b) Contour used in Eq.~(\ref{rho}). Red wavy lines show the correlation function $L(z-z')$ in $F_{\text{FV}}[x,y,\bar{x}]$. (c-d) Contour used in the characteristic function of work. Time-dependences of the external control are different in Eq.~(\ref{WCFone}) and Eq.~(\ref{WCFtwo}).    
}
\label{fig:CFW}
\end{center}
\end{figure}

\noindent{\it Quantum work functional and work statistics in the path integral formalism.---} 
To obtain the path integral expression of Eq.~(\ref{isoCF}), we note the following relations: $\langle x_{f}|U_{\rm{S}}e^{-i\nu H_{\rm{S}}(\lambda_{0})}|x_{i}\rangle=\int Dx \hspace{1mm} e^{(i/\hbar)S_{1}^{\nu}[x]}$ and $\langle y_{i}|U^{\dagger}_{\rm{S}}e^{i\nu H_{\rm{S}}(\lambda_{\tau})}|y_{f}\rangle=\int Dy\hspace{1mm} e^{-(i/\hbar)S_{2}^{\nu}[y]}$, where the actions $S_{1}^{\nu}[x]$ and $S_{2}^{\nu}[y]$ are defined as
\beqa
S_{1}^{\nu}[x]&:=&\int^{\hbar\nu}_{0}dt \mathcal{L}[\lambda_{0},x(t)]+\int^{\tau+\hbar\nu}_{\hbar\nu}\hspace{-1mm}dt \mathcal{L}[\lambda_{t-\hbar\nu},x(t)], \nonumber \\
S_{2}^{\nu}[y]&:=&S[y]+\int^{\tau+\hbar\nu}_{\tau}\hspace{-1mm}ds \mathcal{L}[\lambda_{\tau},y(s)]. 
\label{actionone}
\eeqa
Here, $S[y]:=\int^{\tau}_{0}ds\mathcal{L}[\lambda_{s},y(s)]$ is the usual action and $\mathcal{L}[\lambda_{s},y(s)]:=\frac{M}{2}\dot{y}^{2}(s)-V(\lambda_{s},y(s))$ is the Lagrangian. As a result, we can rewrite Eq.~(\ref{isoCF}) as
\beq
\chi_{W}(\nu)=\int e^{\frac{i}{\hbar}(S_{1}^{\nu}[x]-S_{2}^{\nu}[y])}\rho(x_{i},y_{i}) ,\label{isoWCFone}
\eeq
where $\rho(x_{i},y_{i}):=\langle x_{i}|\rho_{\rm{S}}(0)|y_{i}\rangle$ and the integration in Eq.~(\ref{isoWCFone}) is performed over $\int dx_{i}dy_{i}dx_{f}dy_{f}\delta(x_{f}-y_{f})\int Dx\int Dy$. In Eq.~(\ref{isoWCFone}), the time dependence of the controlling parameter $\lambda_{t}$ between the forward $x(t)$ and the backward $y(s)$ paths are shifted by $\hbar\nu$, which is relevant to the Ramsey interferometry scheme proposed in Ref.~\cite{Ramsey} (see also Fig.~\ref{fig:CFW} (a)). Next, we use the identity $(i/\hbar)S_{1}^{\nu}[x]=(i/\hbar)S_{2}^{\nu}[x]+i\nu W_{\nu}[x]$~\cite{identity} and rewrite Eq.~(\ref{isoWCFone}) as
\beq
\chi_{W}(\nu)=\int e^{\frac{i}{\hbar}(S_{2}^{\nu}[x]-S_{2}^{\nu}[y])} \rho(x_{i},y_{i}) e^{i\nu W_{\nu}[x]}. \label{isoWCFtwo}
\eeq
Here, we introduce the quantum work functional along the forward path $x(t)$~\cite{commenta}
\beq
W_{\nu}[x]:=\int^{\tau}_{0}\hspace{-1mm}dt\hspace{1mm} \frac{1}{\hbar\nu}\int^{\hbar\nu}_{0}\hspace{-2mm} ds  \dot{\lambda}_{t}\frac{\partial V[\lambda_{t},x(t+s)]}{\partial \lambda_{t}} . \label{WFunction}
\eeq
It can be regarded as a quantum extension of the classical work defined as the integral of the supplied power~\cite{Sekimoto,Jarzynski1}
\beq
W_{\text{cl}}[x]:=\int^{\tau}_{0}dt \dot{\lambda}_{t} \frac{\partial V[\lambda_{t},x(t)]}{\partial \lambda_{t} }. \label{CLW}
\eeq 
Both $W_{\nu}[x]$ and $W_{m,n}$ lead to the same work statistics (Eq.~(\ref{isoCF}) and Eq.~(\ref{isoWCFtwo}) are identical). 
In Ref.~\cite{Hanggiaspects}, it is pointed out that the equivalence of power and two-point measurement based work definitions fails to hold in quantum mechanics. We would like to emphasize that their conclusion is due to the fact that they did not obtain the proper quantum extension of Eq.~(\ref{CLW}). 
From Eq.~(\ref{WFunction}), we find that a time-average $(\hbar\nu)^{-1}\int^{\hbar\nu}_{0}ds\cdots$ is required to circumvent the uncertainty relation and obtain the high-frequency component $\nu$ of the work functional. 

By performing the $\hbar$ expansion (or the $\nu$ expansion) in the quantum work functional~(\ref{WFunction}), we can systematically obtain the quantum corrections to the classical expression of the work functional:
\beq
W_{\nu}[x]=W_{\text{cl}}[x]+\frac{i\nu}{2} W_{\text{q}}^{(1)}[x]-\frac{\nu^{2}}{3!}W^{(2)}_{\text{q}}[x]+\cdots , \label{CLWexpansion}
\eeq
where 
\beq
 W_{\text{q}}^{(1)}[x]:=-i\hbar\int^{\tau}_{0}dt \dot{x}(t)\dot{\lambda}_{t}\frac{ \partial^{2}V[\lambda_{t},x(t)]}{\partial \lambda_{t}\partial x(t)} \label{QCW}
\eeq
is the first-order quantum correction to Eq.~(\ref{CLW}). Further quantum corrections can be obtained by Taylor expanding Eq.~(\ref{WFunction}). Using the formula $\langle W^{n}\rangle:=(-i)^{n}\partial^{n}_{\nu}\chi_{W}(\nu)|_{\nu=0}$, we can calculate the moments of work as follows~\cite{commentaa}:
\beq
\langle W^{n}\rangle=\int e^{\frac{i}{\hbar}(S[x]-S[y])}\rho(x_{i},y_{i}) \left. (-i)^{n}\partial^{n}_{\nu}e^{i\nu W_{\nu}[x]}\right|_{\nu=0}. 
 \label{Wfinitemoment}
\eeq
The expansion~(\ref{CLWexpansion}) is useful for calculating the $n$-th moment of work distribution via Eq.~(\ref{Wfinitemoment}). An important observation in this path integral expression is that the quantum corrections to the classical work functional can be found starting from the second moment of work distribution:
\beq
\av{W}=\langle W_{\text{cl}} \rangle_{\text{q-path}},\  \langle W^{2}\rangle=\langle W^{2}_{\text{cl}} \rangle_{\text{q-path}}+\langle W_{\text{q}}^{(1)} \rangle_{\text{q-path}}, \label{Wfirstsecondmoment}
\eeq
where $\langle \bullet \rangle_{\text{q-path}}$ means average over all Feynman quantum paths; $\av{f}_{\text{q-path}}:=\int e^{\frac{i}{\hbar}(S[x]-S[y])}\rho(x_{i},y_{i})f[x]$. 
In general, the $n$-th order quantum correction
appears in the $n+1$-th moment of the work distribution. 

In the semiclassical limit ($\hbar\rightarrow 0$), the quantum work functional~(\ref{WFunction}) reduces to the classical fluctuating work~(\ref{CLW}), and the center coordinate $X(t):=(x(t)+y(t))/2$ behaves as the classical position of the system~\cite{Weiss}. 
By taking the stationary phase approximation, Eq.~(\ref{isoWCFtwo}) converges to its classical counterpart $\langle e^{i\nu W_{\text{cl}}} \rangle_{\text{cl-path}}$~\cite{commentclassicallimit}. Here $\av{f}_{\text{cl-path}}=\int \delta(M\ddot{X}(t)+V'[X(t)]) p(X_{i},\dot{X}_{i})f[X]$ means average over all classical paths obeying Newton's equation, and $p(X_{i},\dot{X}_{i})$ is the initial phase-space distribution. Therefore, we analytically prove the quantum-classical correspondence of the characteristic function of work distribution in isolated systems. Relevant results have been obtained in Refs.~\cite{QCcorrespondence1,QCcorrespondence2} using a different technique. 




\noindent{\it Path integral formalism for an open system.---} Having established a path integral formalism for an isolated system, we generalize it to the open system (see Fig.~\ref{fig:classification}) --quantum Brownian motion described by Caldeira-Leggett model~\cite{Caldeira83,CaldeiraBook}. We use the Caldeira-Leggett model for two reasons. First, the semi-classical limit of this model reproduces the Langevin equation with inertia term~\cite{CaldeiraBook,Caldeira83}, which is a prototype model in the study of classical stochastic thermodynamics~\cite{Sekimoto,Sekimoto98,Seifert,Stochasticbook}. Second, we can analytically integrate out the degrees of freedom of the heat bath, which brings important insights into the understandings of the work statistics in the open quantum system. The Hamiltonian of the composite system is given by $H_{\text{tot}}(\lambda_{t})=H_{\rm{S}}(\lambda_{t})+H_{\rm{B}}+H_{\rm{SB}}$, with
\beqa
& &H_{\rm{S}}(\lambda_{t})=\frac{\hat{p}^{2}}{2M}+\hat{V}(\lambda_{t},\hat{x}), H_{\rm{B}}=\sum_{k}\left(\frac{\hat{p}_{k}^{2}}{2m_{k}}+\frac{m_{k}\omega_{k}^{2}}{2}\hat{q}_{k}^{2}\right), \nonumber \\
& &H_{\rm{SB}}=-\hat{x}\otimes\sum_{k}c_{k}\hat{q}_{k}+\sum_{k}\frac{c^{2}_{k}}{2m_{k}\omega_{k}^{2}}\hat{x}^{2}, \label{Hamiltonian}
\eeqa
where we have included the counter term $\sum_{k}(c_{k}^{2}/2m_{k}\omega_{k}^{2})\hat{x}^{2}$ in the interaction Hamiltonian to cancel the negative frequency shift of the potential~\cite{Caldeira83b}.
Here $H_{\rm{S}}(\lambda_{t})$ is the same Hamiltonian we use for an isolated system, and $m_{k}$, $\omega_{k}$, $c_{k}$, $\hat{q}_{k}$ and $\hat{p}_{k}$ are the mass, frequency, coupling strength, position and momentum of the $k$-th mode of the bath, respectively. 

The reduced density matrix of the system at time $\tau$ is given by $\rho_{\rm{S}}(\tau)=\Tr_{\rm{B}}[U_{\rm{SB}}\rho(0)U^{\dagger}_{\rm{SB}}]$, where $U_{\rm{SB}}=\hat{\text{T}}[\exp(-\frac{i}{\hbar}\int^{\tau}_{0}dt H_{\text{tot}}(\lambda_{t}))]$ is the unitary time-evolution operator for the composite system and we choose the initial state to be $\rho(0)=\exp(-\beta H_{\text{tot}}(\lambda_{0}))/Z_{\text{tot}}(\lambda_{0})$. Using the path-integral technique, the reduced density matrix takes the form~\cite{Caldeira87,Weiss,Grabert,Tanimura}
\beqa
& &\langle x_{f}|\rho_{\rm{S}}(\tau)| y_{f}\rangle= Z_{\lambda_{0}}^{-1}\int dx_{i}dy_{i}\int^{x(\tau)=x_{f}}_{x(0)=x_{i}}\hspace{-1mm} Dx \int^{y(\tau)=y_{f}}_{y(0)=y_{i}}\hspace{-1mm} Dy  \nonumber \\
& &\times\int^{\bar{x}(\hbar\beta)=x_{i}}_{\bar{x}(0)=y_{i}}D\bar{x}\     e^{\frac{i}{\hbar}(S[x]-S[y])-\frac{1}{\hbar}S^{\rm{E}}[\bar{x}]}F_{\text{FV}}[x,y,\bar{x}] , \label{rho}
\eeqa
where $F_{\text{FV}}[x,y,\bar{x}]$ is the generalized Feynman-Vernon influence functional~\cite{Caldeira87,Grabert}, and $x$, $y$, $\bar{x}$ are the forward, backward, imaginary time coordinates of the system, respectively (see also the contour $\mathcal{C}_{1}$ in Fig.~\ref{fig:CFW} (b)). Here, $S[x]$ is the action, $S^{\rm{E}}[\bar{x}]:=\int^{\hbar\beta}_{0}du(\frac{M}{2}\dot{\bar{x}}^{2}(u)+V[\lambda_{0},\bar{x}(u)])$ is the Euclidian version of the action, and $Z_{\lambda_{0}}:=
\Tr[e^{-\beta H_{\text{tot}}(\lambda_{0})}]/\Tr[e^{-\beta H_{\rm{B}}}]$ is the reduced partition function of the system.


\noindent{\it Work statistics for the Caldeira-Leggett model.---} 
By generalizing Eq.~(\ref{isoCF}) to the case of the composite system, the characteristic function of work distribution is given by
\beq
\chi_{W}(\nu)=\Tr\left[ U_{\rm{SB}}e^{-i\nu H_{\text{tot}}(\lambda_{0})}\rho(0)U^{\dagger}_{\rm{SB}} e^{i\nu H_{\text{tot}}(\lambda_{\tau})}\right]. \label{WCF}
\eeq
We can integrate out the bath degrees of freedom and obtain the path integral expression of Eq.~(\ref{WCF}) by adapting a similar technique we use for the isolated system:
\beq
\chi_{W}(\nu) =Z^{-1}_{\lambda_{0}}\int e^{\frac{i}{\hbar}(S_{1}^{\nu}[x]-S_{2}^{\nu}[y])-\frac{1}{\hbar}S^{\rm{E}}[\bar{x}]} F_{\text{FV}}^{\nu}[x,y,\bar{x}]. \label{WCFone}
\eeq
Here, the integration  is performed over $\int \delta(x_{f}-y_{f})dx_{i}dy_{i}dx_{f}dy_{f}DxDyD\bar{x}$ and the influence functional is given by 
\begin{widetext}
\beqa
& &F_{\rm{FV}}^{\nu}[x,y,\bar{x}] =\exp\Bigl[ -\frac{1}{\hbar}\int^{\tau+\hbar\nu}_{0}\hspace{-1mm}dt\int^{t}_{0}ds (x(t)-y(t))(L(t-s)x(s)-L^{*}(t-s)y(s)) +\frac{i\mu}{\hbar}\int^{\tau+\hbar\nu}_{0}\hspace{-1mm}dt (x^{2}(t)-y^{2}(t)) \nonumber \\ 
& &+\frac{i}{\hbar}\int^{\tau+\hbar\nu}_{0}\hspace{-1mm}dt\int^{\hbar\beta}_{0}du(x(t)-y(t))L^{*}(t-iu)\bar{x}(u) +\frac{1}{\hbar}\int^{\hbar\beta}_{0}\hspace{-1mm}du\int^{u}_{0}\hspace{-1mm}du' L(-iu+iu')\bar{x}(u)\bar{x}(u') -\frac{\mu}{\hbar}\int^{\hbar\beta}_{0}du\bar{x}^{2}(u)\Bigr], \label{nuFV}
\eeqa
\end{widetext}
where $L(t-iu):=\sum_{k}\frac{c_{k}^{2}}{2m_{k}\omega_{k}}(\cosh\frac{\hbar\omega_{k}\beta}{2}\cosh\omega_{k}(u+i t)-\sinh \omega_{k}(u+it))$ is the complex bath correlation function, and $\mu:=\sum_{k}c^{2}_{k}/(2m_{k}\omega_{k}^{2})$. See Fig.~\ref{fig:CFW} (c) for the contour $\mathcal{C}_{2}$ we use in Eq.~(\ref{WCFone}). Also note that by taking $\nu=0$, Eq.~(\ref{nuFV}) reproduces $F_{\text{FV}}[x,y,\bar{x}]$. The actions $S_{1}^{\nu}[x]$ and $S_{2}^{\nu}[y]$ are the same as that we use for the isolated system~(\ref{actionone}). 
Using again the identity~\cite{identity}, the path integral expression of the characteristic function of work distribution for an open system is given by (see Fig.~\ref{fig:CFW} (d) for the contour)
\beqa
\hspace{-2mm}\chi_{W}(\nu) &=&Z^{-1}_{\lambda_{0}}\int dx_{f}dy_{f}dx_{i}dy_{i}\delta(x_{f}-y_{f})\int DxDyD\bar{x}  \nonumber \\
&\times&\hspace{-0.5mm} e^{\frac{i}{\hbar}(S_{2}^{\nu}[x]-S_{2}^{\nu}[y])-\frac{1}{\hbar}S^{\rm{E}}[\bar{x}]}F_{\text{FV}}^{\nu}[x,y,\bar{x}]e^{i\nu W_{\nu}[x]}, \label{WCFtwo}
\eeqa
where the quantum work functional is given by Eq.~(\ref{WFunction}). 
We note that Eq.~(\ref{WCFtwo}) is valid for the strong-coupling, non-Markovian, and non-RWA regime, and it allows us to calculate work statistics of the quantum Brownian model. The moments of work can be calculated by 
using Eq.~(\ref{Wfinitemoment}), but the average is over all Feynman paths for the open system dynamics~(\ref{rho}): $\langle f\rangle_{\text{q-path}}=Z_{\lambda_{0}}^{-1}\int e^{\frac{i}{\hbar}(S[x]-S[y])-\frac{1}{\hbar}S^{\rm{E}}[\bar{x}]}F_{\text{FV}}[x,y,\bar{x}]f[x]$. In particular, Eq.~(\ref{Wfirstsecondmoment}) also holds for an open system using the above path integral average. We can show the Jarzynski equality using the path integral expression by using Eq.~(\ref{WCFtwo})~\cite{footnoteJar}. 

To show the quantum-classical correspondence of the characteristic function of work in the Brownian motion model,  we take $\hbar\rightarrow 0$ and $\beta\rightarrow 0$ and introduce $X(t)=(x(t)+y(t))/2$. We follow the standard treatment~\cite{Weiss,Tanimura} to obtain the quasiclassical (non-Markovian) Langevin equation by introducing the noise function $\Omega(t):=i\int^{\tau}_{0}ds(x(s)-y(s))\text{Re}[L(t-s)]$. 
Using a method similar to that of the isolated system, we prove that in the  classical limit, Eq.~(\ref{WCFtwo}) converges to its classical counterpart $\langle e^{i\nu W_{\text{cl}}}\rangle_{\text{cl-path}}$~\cite{commentclassicallimit}. Here, $\langle f\rangle_{\text{cl-path}}$ 
is the average over all classical paths satisfying the non-Markovian Langevin equation $M\ddot{X}(t)+V'[X(t)]+\int^{t}_{0}dsK(t-s)\dot{X}(s)=\Omega(t)$, where $K(t):=\sum_{k}(c_{k}^{2}/m_{k}\omega_{k}^{2})\cos\omega_{k}t$
 is the classical bath-correlation function. We emphasize that the introduction of the work functional along individual Feynman path enables us for the first time to show the quantum-classical correspondence of the work statistics in open systems (see Fig.~\ref{fig:classification}).



\noindent{\it Example: dragged harmonic oscillator.---} In order to demonstrate the effectiveness of our approach in calculating work statistics, let us consider a potential given by $V[\lambda_{t},x(t)]=\frac{M\omega^{2}}{2}(x(t)-\lambda_{t})^{2}$. Here, $\lambda_{t}$ describes the time dependence of the center of the harmonic potential, and we consider a linear protocol $\lambda_{t}=vt$. We note that the characteristic function of work for an isolated system is analytically calculated in Ref.~\cite{Talkner08} by utilizing the concept of work based on two-point measurement. In Ref.~\cite{Supplement}, we obtain the same result by using our path integral approach. 

\begin{figure}[t]
\begin{center}
\includegraphics[width=.48\textwidth]{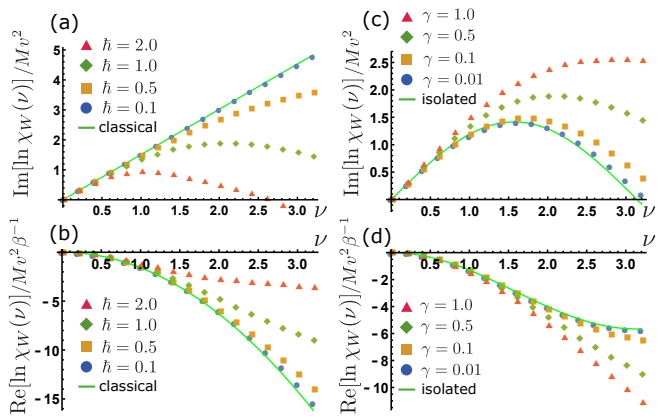}
\caption{ Plot of the characteristic function of work distribution for a dragged harmonic oscillator using the Caldeira-Leggett model~(\ref{Hamiltonian}). The analytical expression of $\chi_{W}(\nu)$ is given in Ref.~\cite{supplement}. For simplicity, we choose the high-temperature regime and choose the following parameters: $M=\omega=v=1$, $\tau=2$, $\beta=0.01$. 
We choose the Ohmic spectrum $J(\omega):=\sum_{k}(\pi c_{k}^{2}/2m_{k}\omega_{k})\delta(\omega-\omega_{k})=M\gamma\omega$ with a high-frequency cutoff $\omega_{\mathrm{D}}$, where $\gamma$ is the friction coefficient. 
(a-b) Plot of $\text{Im}[\ln\chi_{W}(\nu)]$ and $\text{Re}[\ln\chi_{W}(\nu)]$ for different values of $\hbar$ (we set $\gamma=0.5$). 
(c-d) Plot of $\text{Im}[\ln\chi_{W}(\nu)]$ and $\text{Re}[\ln\chi_{W}(\nu)]$ for different values of $\gamma$ (we set $\hbar=1$).}
\label{fig:CFW2}
\end{center}
\end{figure}

For an open system described by the Caldeira-Leggett model~(\ref{Hamiltonian}), we cannot apply the two-point measurement approach in practice because of the huge number of degrees of freedom of the bath. However, the introduction of the work functional~(\ref{WFunction}) enables us to analytically calculate the characteristic function of work distribution~(\ref{WCFtwo}) by using techniques~\cite{Caldeira90} developed in the field of path integral for open quantum systems~\cite{Supplement}. We plot $\chi_{W}(\nu)$ in Fig.~\ref{fig:CFW2}.   

\noindent{\it Summary.---} 
Before concluding the paper, we would like to give the following remarks. The usual two-point measurement based quantum work is good for demonstrating Jarzynski equality~\cite{Campisi09} but practically cannot be used to studying work statistics in an open quantum system because we have to deal with a huge number of degrees of freedom of the bath. By contrast, with the work functional along individual Feynman path, we can not only demonstrate the Jarzynski equality but we can also calculate the work statistics and show the convergence of the quantum work statistics to its classical counterpart. In addition, the work functional along individual Feynman path provides important insights into our understandings about work in quantum systems. Thus, the path integral approach to quantum work has both conceptual and technical advantages over the two-point measurement approach to quantum work. 

In this Letter, we invented a path integral approach to study the quantum work and its statistics in the non-Markovian, non-RWA, and strong coupling regime using the quantum Brownian motion model. 
In comparison with the definition of work based on two-point measurement, the work functional along individual Feynman path~(\ref{WFunction}) introduced in our paper offers conceptually different interpretations and physical intuitions about work in quantum systems.
 Through the $\hbar$ expansion, we can systematically obtain quantum corrections to the classical work. In the strong-coupling quantum Brownian model, this work functional enables us to calculate the work statistics 
and prove analytically the quantum-classical correspondence of both the work functional and the work statistics, which has not been reported in open systems so far. 
In addition, we use a dragged harmonic oscillator as an example to show the corrections and the convergence of the quantum work statistics to its classical counterpart in an open quantum system. 

\begin{acknowledgments}
The authors thank Prof. Christopher Jarzynski, Prof. Amir Ordacgi Caldeira, Prof. Erik Aurell and Prof. Peter H\"{a}nggi for helpful discussions and comments. This work was supported by the National Science Foundation of China under Grants No.~11775001, 11375012 and 11534002, and The Recruitment
 Program of Global Youth Experts of China.


\end{acknowledgments}

\widetext
\clearpage
\begin{center}
\textbf{\large Supplemental Material: Path integral approach to quantum thermodynamics}\\
\vspace{4mm}

Ken Fun$\text{o}^{1}$ and H. T. Qua$\text{n}^{1,2,*}$\\
\vspace{2mm}

{\it {\small $^{1}$School of Physics, Peking University, Beijing 100871, China}} \\
{\it {\small $^{2}$Collaborative Innovation Center of Quantum Matter, Beijing 100871, China}}
\end{center}
\setcounter{equation}{0}
\setcounter{figure}{0}
\setcounter{table}{0}
\setcounter{page}{1}
\makeatletter
\renewcommand{\theequation}{S\arabic{equation}}
\renewcommand{\thefigure}{S\arabic{figure}}
\renewcommand{\bibnumfmt}[1]{[S#1]}
\renewcommand{\citenumfont}[1]{S#1}

In this supplementary material, we give a detailed derivation of the classical limit of the characteristic function of work in Sec.~\ref{sec:cl}. In Sec.~\ref{sec:supp:Jar}, we derive Jarzynski's equality based on the path integral expression~(16). In Sec.~\ref{sec:ex}, we analytically calculate the characteristic function of work for a dragged harmonic oscillator.

\section{\label{sec:cl}Convergence of the quantum characteristic function of work distribution to its classical counterpart}
In this section, we show the convergence of the quantum characteristic function of work distribution to its classical counterpart in detail. 
\subsection{Isolated system case}
We first take the lowest order $\hbar$ terms for the action of the forward and backward paths, and obtain
\beq
\frac{i}{\hbar}S^{\nu}_{2}[x]-\frac{i}{\hbar}S^{\nu}_{2}[y]=-\frac{i}{\hbar}\int^{\tau}_{0}dt \xi(t)\Bigl( M\ddot{X}(t)+V'(X) \Bigr)-\frac{i}{\hbar}M\xi(0)\dot{X}(0)+ O(\xi^{3})+O(\hbar) \label{supp:SXY}.
\eeq 
Here, we define $X=(x+y)/2$ and $\xi=x-y$, and we expand the potential energy as $V(x)-V(y)=V(X+\xi/2)-V(X-\xi/2)=\xi V'(X)+O(\xi^{3})$ in Eq.~(\ref{supp:SXY}). Because of the delta function $\delta(x_{f}-y_{f})$ in the characteristic function of work, we can set $\xi(\tau)=x_{f}-y_{f}=0$ and thus $O(\hbar^{0})$ terms vanish. 
Next, we use the notation $\dot{X}_{i}=\dot{X}(0)$ and introduce the Wigner function
\beq
p(X_{i},\dot{X}_{i}):=\frac{1}{\pi\hbar}\int d\xi_{i}e^{-(i/\hbar)M\xi_{i}\dot{X}_{i}}\rho(X_{i},\xi_{i}), \label{supp:Wigner}
\eeq
which converges to the classical phase-space distribution in the $\hbar\rightarrow 0$ limit. Using Eqs.~(\ref{supp:SXY}) and (\ref{supp:Wigner}) and keeping the lowest order $\hbar$ terms, the characteristic function of work distribution~(4) takes the form
\beq
\chi_{W}(\nu)=\pi\hbar \int dX_{f} dX_{i} \int DX \int D\xi  \ e^{ -(i/\hbar)\int^{\tau}_{0}dt\xi(t)(M\ddot{X}(t)+V'[X(t)])}p(X_{i},\dot{X}_{i})e^{i\nu W_{\text{cl}}[X]}+O(\hbar).
\eeq  
Integration over $D\xi$ gives a delta function $\delta(M\ddot{X}(t)+V'[X(t)])$, and we show the quantum-classical correspondence of the work statistics for an isolated system:
\beq
\chi_{W}(\nu)=\int dX_{f} dX_{i} \int DX  \delta(M\ddot{X}(t)+V'[X(t)]) e^{i\nu W_{\text{cl}}[X]}p(X_{i},\dot{X}_{i})+O(\hbar)=\av{e^{i\nu W_{\text{cl}}}}_{\text{cl-path}}+O(\hbar).
\eeq

\subsection{Open system case}
Let us consider the lowest order $\hbar$ expansion in the generalized Feynman-Vernon influence functional~(15):
\beqa
F^{\nu}_{\text{FV}}[x,y,\bar{x}]&=&\exp\biggl[ -\frac{1}{\hbar}\int^{\tau}_{0}dt\int^{t}_{0}ds (x(t)-y(t))(L(t-s)x(s)-L^{*}(t-s)y(s))  +\frac{i\mu}{\hbar}\int^{\tau}_{0}dt (x^{2}(t)-y^{2}(t)) \nonumber \\
& &\ \ \ \ +\frac{i}{\hbar}\int^{\tau}_{0}dt \int^{\hbar\beta}_{0}du L^{*}(t-iu) \Bigl(x(t)-y(t)\Bigr) \bar{x}(u) \nonumber \\
& &\ \ \ \ +\frac{1}{\hbar}\int^{\hbar\beta}_{0}du\int^{u}_{0}du' L(-iu+iu')\bar{x}(u)\bar{x}(u')-\frac{\mu}{\hbar}\int^{\hbar\beta}_{0}du \bar{x}^{2}(u) +O(\hbar) \biggr]. \label{Supp:FV}
\eeqa
Note that $O(\hbar^{0})$ terms vanish because  $\xi(\tau)=0$. The first two terms inside the exponential can be calculated as
\beqa
& &-\frac{1}{\hbar}\int^{\tau}_{0}dt\int^{t}_{0}ds (x(t)-y(t))(L(t-s)x(s)-L^{*}(t-s)y(s))  +\frac{i\mu}{\hbar}\int^{\tau}_{0}dt (x^{2}(t)-y^{2}(t)) \nonumber \\
&=& -\frac{1}{2\hbar}\int^{\tau}_{0}dt\int^{\tau}_{0}ds \xi(t)L_{\text{Re}}(t-s)\xi(s)-\frac{i}{\hbar}\int^{\tau}_{0}dt\int^{t}_{0}ds  K(t-s) \xi(t)\dot{X}(s)-\frac{i}{\hbar}X(0)\int^{\tau}_{0}dt K(t)\xi(t), \label{supp:imaginaryxx}
\eeqa
where $L_{\text{Re}}(t):=\text{Re}[L(t)]$ and $K(t)=\sum_{k}(c^{2}_{k}/m_{k}\omega^{2}_{k})\cos\omega_{k}t$ is the classical bath correlation function. By taking the high-temperature limit for the second line in Eq.~(\ref{Supp:FV}), we have
\beqa
& &\frac{i}{\hbar}\int^{\tau}_{0}dt \int^{\hbar\beta}_{0}du L^{*}(t-iu) \Bigl(x(t)-y(t)\Bigr) \bar{x}(u) \nonumber \\
&=&\frac{i}{\hbar}\int^{\tau}_{0}dt \int^{\hbar\beta}_{0}du \Bigl(x(t)-y(t)\Bigr)\bar{x}(u) \sum_{k}\frac{c_{k}^{2}}{2m_{k}\omega_{k}}\frac{\sinh(\hbar\omega_{k}\beta/2-\omega_{k} u)\sinh(i\omega_{k}t )}{\sinh\hbar\omega_{k}\beta/2} \nonumber \\
& &-\frac{i}{\hbar}\int^{\tau}_{0}dt \Bigl(x(t)-y(t)\Bigr) \sum_{k}\frac{c_{k}^{2}}{2m_{k}\omega_{k}^{2}} \cosh(i\omega_{k}t) \biggl\{ \Bigl[ \frac{\sinh(\hbar\omega_{k}\beta/2-\omega_{k}u)}{\sinh(\hbar\omega\beta/2)}\bar{x}(u)\Bigl]^{\hbar\beta}_{0}-\int^{\hbar\beta}_{0}du \dot{\bar{x}}(u)\frac{\sinh(\hbar\omega_{k}\beta/2-\omega_{k}u)}{\sinh(\hbar\omega\beta/2)} \biggr\} \nonumber \\
&=&\frac{i}{\hbar}X(0)\int^{\tau}_{0}dt K(t)\xi(t) +O(\beta). \label{supp:bathsystemcal}
\eeqa

We combine Eqs.~(\ref{supp:SXY}) and (\ref{Supp:FV}-\ref{supp:bathsystemcal}) and obtain the characteristic function of work (16) in the $\hbar\rightarrow 0$ and $\beta\rightarrow 0$ limit:
\beqa
\chi_{W}(\nu)&=&\int dx_{i}dy_{i}dx_{f}dy_{f}\delta(x_{f}-y_{f}) Dx Dy  D\bar{x} e^{-\frac{1}{\hbar}S^{(\text{E})}[\bar{x}]+\frac{i}{\hbar}(S^{\nu}_{2}[x]-S^{\nu}_{2}[y])}F_{\text{FV}}^{\nu}[x,y,\bar{x}] e^{i\nu W_{\nu}[x]}\nonumber \\
&=&\int dX_{f}dX_{i}\int d\xi_{i} \int DX \int D\xi \int D\Omega P[\Omega]  e^{-\frac{i}{\hbar}M\dot{X}_{i}\xi_{i}}\rho(X_{i},\xi_{i})e^{i\nu W_{\text{cl}}[X]} \nonumber \\
& &\times \exp\biggl[ -\frac{i}{\hbar}\int^{\tau}_{0}dt \xi(t)\Bigl( M\ddot{X}(t)+V'[X(t)] +\int^{t}_{0}ds K(t-s)\dot{X}(s)-\Omega(t)\Bigr) \biggr]+O(\hbar,\beta), \label{supp:sysfinal}
\eeqa
where the reduced canonical distribution of the system is given by 
\beq
\rho(X_{i},\xi_{i})=\frac{1}{Z_{\lambda_{0}}}\int D\bar{x}  \exp\Bigl(-\frac{1}{\hbar} S^{(\text{E})}[\bar{x}] +\frac{1}{\hbar}\int^{\hbar\beta}_{0}du \int^{u}_{0}du' L(-iu+iu')\bar{x}(u)\bar{x}(u') -\frac{\mu}{\hbar}\int^{\hbar\beta}_{0}du \bar{x}^{2}(u)  \Bigr), \label{supp:reducedrho}
\eeq
and we introduce the noise
\beqa
\Omega(t):=i\int^{\tau}_{0}ds L_{\text{Re}}(t-s)\xi(s) , \label{supp:NOISE}
\eeqa
and the weight function
\beq
P[\Omega]=C^{-1}\exp\biggl[ -\frac{1}{2\hbar}\int^{\tau}_{0}dt\int^{\tau}_{0}ds \Omega(t)L^{-1}_{\text{Re}}(t-s)\Omega(s) \biggr],
\eeq
with $C$ being the normalization constant. 
By taking the high-temperature (classical) limit, we have $L_{\text{Re}}(t)=(1/\hbar\beta)K(t)+O(\beta)$. Therefore, the noise $\Omega(s)$ satisfies the classical properties in the high-temperature limit:
\beqa
& &\av{\Omega(t)}=0,\\
& &\av{\Omega(t)\Omega(s)}=\hbar L_{\text{Re}}(t-s)=\beta^{-1}K(t-s)+O(\beta).
\eeqa
We introduce the Wigner function by Eq.~(\ref{supp:Wigner}) and integrate over $D\xi$ to finally obtain
\beq
\chi_{W}(\nu)=\int dX_{f}dX_{i} \int DX  \int D\Omega P[\Omega] \delta\Bigl( M\ddot{X}(t)+V'[X(t)] +\int^{t}_{0}ds K(t-s)\dot{X}(s)-\Omega(t)\Bigr) p(X_{i},\dot{X}_{i})e^{i\nu W_{\text{cl}}[X]}+O(\hbar,\beta), \label{supp:classicallimit}
\eeq
where the delta function enforces the classical path $X(t)$ to satisfy the classical non-Markovian Langevin equation:
\beq
M\ddot{X}(t)+V'[X(t)] +\int^{t}_{0}ds K(t-s)\dot{X}(s)=\Omega(t).
\eeq
Equation~(\ref{supp:classicallimit}) is the classical characteristic function of work and thus we show the quantum characteristic function of work distribution converges to its classical counterpart.

\section{\label{sec:supp:Jar}Jarzynski's equality} 
 Jarzynski's equality can be shown by taking $\nu=i\beta$ in the characteristic function of work~\cite{characteristic1,Campisi09}. From Eq.~(13), we have
\beq
\chi_{W}(i\beta)=\int dW e^{-\beta W}P(W)=\av{e^{-\beta W}}=e^{-\beta\Delta F}.\label{Jarzynski}
\eeq
Here, $\Delta F:=F_{\lambda_{\tau}}-F_{\lambda_{0}}$, where $F_{\lambda_{t}}
:=-\beta^{-1}\ln Z_{\lambda_{t}}$ is the free energy of the open system of interest~\cite{Campisi09}. We can also show the Jarzynski equality using the path integral expression by using Eq.~(16). We note that taking $\nu=i\beta$ requires a Wick rotation, and the quantum work functional~(5) can be expressed as
\beq
-\beta W_{\beta}[\{\bar{x}_{t}\}]=-\int^{\tau}_{0}dt\dot{\lambda}_{t}\frac{\partial}{\partial \lambda_{t}}\frac{1}{\hbar}S^{\rm{E}}[\lambda_{t},\bar{x}_{t}],
\eeq
where $S^{\rm{E}}[\lambda_{t},\bar{x}_{t}]=\int^{\hbar\beta}_{0}du[M\dot{\bar{x}}_{t}^{2}(u)/2+V(\lambda_{t},\bar{x}_{t}(u))]$ with endpoint conditions $\bar{x}_{t}(0)=x(t)$ and $\bar{x}_{t}(\hbar\beta)=y(t)$. Then, we find that 
\beq
F^{i\beta}_{\text{FV}}[x,y,\bar{x}]e^{-\beta W_{\beta}[\{\bar{x}_{t}\}]}=\tilde{F}_{\text{FV}}[x,y,\bar{x}_{\tau}],
\eeq
where $\tilde{F}_{\text{FV}}$ is calculated from the time-reversal of the contour $\mathcal{C}_{1}$ (Fig.~2 (b)). This gives a density matrix $\tilde{\rho}^{S}(\tau)$ generated from the time-reversed protocol. Therefore, 
\beq
\chi_{W}(i\beta)=\Tr[\tilde{\rho}_{\rm{S}}(\tau)]e^{-\beta\Delta F}=e^{-\beta\Delta F}
\eeq
and this is the Jarzynski's equality.


\section{\label{sec:ex}Characteristic function of work for a dragged harmonic oscillator}
In this section, as an example, we obtain analytical results for a dragged harmonic oscillator. The potential is given by $V[\lambda_{t},x(t)]=\frac{M\omega^{2}}{2}(x(t)-\lambda_{t})^{2}$, where we choose a linear dragging protocol $\lambda_{t}=vt$. 

\subsection{Isolated system}
 The characteristic function of work for an isolated system is analytically calculated in Ref.~\cite{Talkner08} and it takes the form 
\beq
\chi_{W}(\nu)=\exp \Bigl[ \frac{M\omega}{\hbar}\Bigl( i\sin\hbar\omega\nu-(1-\cos\hbar\omega\nu)\coth\frac{\hbar\omega\beta}{2}\Bigr)f(\tau)\Bigr],
\eeq 
where $f(\tau):=\int^{\tau}_{0}dt\int^{t}_{0}ds \cos\omega(t-s)\dot{\lambda}_{t}\dot{\lambda}_{s}.$ The work functional is given by 
\beq
W_{\nu}[x]=\int^{\tau}_{0}dt\frac{1}{\hbar\nu}\int^{\hbar\nu}_{0}ds \dot{\lambda}_{t}M\omega^{2}(\lambda_{t}-x(t+s)).
\eeq
We also note that 
\beqa
W_{\text{cl}}[x]&=&M\omega^{2}\int^{\tau}_{0}dt \dot{\lambda}_{t}(\lambda_{t}-x(t)), \label{supp:CLW}\\
W_{\text{q}}^{(1)}[x]&=&i\hbar M\omega^{2}\int^{\tau}_{0}dt\dot{\lambda}_{t}\dot{x}(t), \\
W_{\text{q}}^{(2)}[x]&=&\hbar^{2}M\omega^{2}\int^{\tau}_{0}dt\dot{\lambda}_{t}\ddot{x}(t).
\eeqa
If we only use the classical expression of work~(\ref{supp:CLW}) for the calculation of the work statistics, we have 
\beq
\langle e^{i\nu W_{\text{cl}}}\rangle_{\text{q-path}}=\exp\Bigl[M\omega^{2}\Bigl\{\Bigl(i\nu-\frac{\nu^{2}\hbar\omega}{2}\coth\frac{\hbar\omega\beta}{2}\Bigr)f(\tau)+\frac{i\hbar\nu^{2}\omega}{2}g(\tau)\Bigr\}\Bigr],
\eeq
where $g(\tau):=\int^{\tau}_{0}dt\int^{t}_{0}ds \sin\omega(t-s)\dot{\lambda}_{t}\dot{\lambda}_{s}$. It gives the correct first moment of work distribution 
\beq
\langle W_{\text{cl}}\rangle_{\text{q-path}}=M\omega^{2}f(\tau),
\eeq
but we find a deviation already in the second moment.  To obtain the correct second moment, we need to take into account the first order quantum correction $W^{(1)}_{\text{q}}[x]$ as in Eq.~(10). We find $\langle W^{(1)}_{\text{q}}\rangle_{\text{q-path}}=i\hbar M\omega^{3}g(\tau)$, and thus 
\beq
\langle W^{2}_{\text{cl}}\rangle_{\text{q-path}}+\langle W^{(1)}_{\text{q}}\rangle_{\text{q-path}}=M^{2}\omega^{4}f^{2}(\tau)+\hbar M\omega^{3}\coth\frac{\hbar\omega\beta}{2}f(\tau)
\eeq
gives the second moment $\av{W^{2}}$ calculated from $\chi_{W}(\nu)$. We further check the validity of the $\hbar$ (or the $\nu$) expansion~(6) up to the second order and find that 
\beq
\langle e^{i\nu W_{\text{cl}}-\frac{\nu^{2}}{2}W^{(1)}_{\text{q}}-\frac{i\nu^{3}}{3!}W^{(2)}_{\text{q}}}\rangle_{\text{q-path}}=\exp\Bigl[M\omega^{2}\Bigl(i\nu-\frac{\hbar\nu^{2}\omega}{2}\coth\frac{\hbar\omega\beta}{2}-\frac{i\hbar^{2}\nu^{3}\omega^{2}}{6}\Bigr)f(\tau)+O(\nu^{4})\Bigr].
\eeq
This is consistent with the exact expression $\chi_{W}(\nu)$ up to $\nu^{3}$ terms. We finally note that the classical characteristic function of work is given by 
\beq
\langle e^{i\nu W_{\text{cl}}}\rangle_{\text{cl-path}}=\exp[M\omega^{2}(i\nu-\nu^{2}\beta^{-1})].
\eeq

\subsection{Open system}

Next, we consider the case of an open system. We note that if we want to calculate the characteristic function in this setup, starting from Eq. (14) is convenient. Our starting point is the following expression for the characteristic function of work [Eq.~(14)]:
\beqa
\chi_{W}(\nu)&=&I_{0}\int dx_{i}dy_{i}dx_{f}dy_{f}\delta(x_{f}-y_{f})\int DxDyD\bar{x} e^{-\frac{1}{\hbar}\bar{S}_{\rm{eff}}+\frac{i}{\hbar}(S^{\nu}_{1}[x]-S^{\nu}_{2}[y])} \nonumber \\
& &\times \exp\biggl[ -\frac{1}{2\hbar}\int^{\tau+\hbar\nu}_{0}dt\int^{\tau}_{0}ds \xi(t)L_{\text{Re}}(t-s)\xi(s)-\frac{iM\gamma}{\hbar}\int^{\tau+\hbar\nu}_{0}dt \xi(t)\dot{X}(t)-\frac{iM\gamma}{\hbar} X_{i}\xi_{i}  \biggr], \label{supp:chiW}
\eeqa
where the effective action $\bar{S}_{\rm{eff}}$ is given by~\cite{Caldeira90}
\beqa
e^{-\frac{1}{\hbar}\bar{S}_{\rm{eff}}}&=&\int D\bar{x} \exp\biggl[ -\frac{1}{\hbar}S^{\rm{E}}[\bar{x}]  +\frac{i}{\hbar}\int^{\tau+\hbar\nu}_{0}\hspace{-1mm}dt \int^{\hbar\beta}_{0}du L^{*}(t-iu) \xi(t) \bar{x}(u) \nonumber \\
& &\ \ \ +\frac{1}{\hbar}\int^{\hbar\beta}_{0}du\int^{u}_{0}du' L(-iu+iu')\bar{x}(u)\bar{x}(u')-\frac{\mu}{\hbar}\int^{\hbar\beta}_{0}du \bar{x}^{2}(u) \biggl] \nonumber \\
&=&I_{0}\exp\biggl[ -\frac{M\gamma}{2\hbar\pi}\int^{\omega_{D}}_{0}d\Omega \frac{\Omega^{3}\coth\frac{\hbar\Omega\beta}{2}}{(\omega^{2}-\Omega^{2})^{2}+\gamma^{2}\Omega^{2}}\left\{ \xi_{i}^{2}-\xi_{i}\int^{\tau+\hbar\nu}_{0}\hspace{-1mm}dt\xi(t)\left( \frac{2(\Omega^{2}-\omega^{2})}{\Omega}\sin\Omega t-2\gamma\cos\Omega t\right)\right\}  \label{effectiveaction}\\
& &-\frac{M}{2\kappa\hbar}\left\{ X_{i}+\frac{i\gamma}{\pi}\int^{\tau+\hbar\nu}_{0}\hspace{-1mm}dt\int^{\omega_{D}}_{0}d\Omega \frac{\Omega\coth\frac{\hbar\beta\Omega}{2}}{(\omega^{2}-\Omega^{2})^{2}+\gamma^{2}\Omega^{2}}\xi(t)\left( (\omega^{2}-\Omega^{2})\cos\Omega t-\gamma\Omega\sin\Omega t\right)\right\}^{2}+\frac{iM\gamma}{\hbar} X_{i}\xi_{i} \biggl]. \nonumber 
\eeqa
Here, $\kappa=\sum_{n=-\infty}^{\infty}(\omega_{n}^{2}+\gamma|\omega_{n}|+\omega^{2})^{-1}$ with $\omega_{n}=2n\pi/(\hbar\beta)$, and $I_{0}$ comes from the Gaussian integral performed in the first line of Eq.~(\ref{effectiveaction}). We choose 
the Ohmic spectrum $J(\omega)=M\gamma\omega$  with a high frequency cutoff $\omega_{\mathrm{D}}$. The classical bath correlation function satisfies $K(t-s)=M\gamma\delta(t-s)$. 
Now Eq.~(\ref{supp:chiW}) takes the form
\beqa
\chi_{W}(\nu)&=&I_{0}\int dx_{i}dy_{i}dx_{f}dy_{f}\delta(x_{f}-y_{f})\int DxDy \exp\biggl[ -\frac{i}{\hbar}\int^{\tau+\hbar\nu}_{0}\hspace{-1mm}dt MX(t)\Bigl( \ddot{\xi}(t)-\gamma\dot{\xi}(t)+\omega^{2}\xi(t)-\omega^{2}(\lambda_{1}(t)-\lambda_{2}(t))\Bigr)  \nonumber \\
& &+\frac{i}{\hbar}MX_{f}\dot{\xi}(\tau+\hbar\nu)-\frac{i}{\hbar}MX_{i}\dot{\xi}(0)+\frac{i}{\hbar}M\gamma X_{i}\xi_{i} +\frac{iM\omega^{2}}{2\hbar}\int^{\tau+\hbar\nu}_{0}\hspace{-1mm}dt (\xi(t)-\lambda_{1}(t)+\lambda_{2}(t))(\lambda_{1}(t)+\lambda_{2}(t)) \nonumber \\
& &-\frac{1}{2\hbar}\int^{\tau+\hbar\nu}_{0}\hspace{-1mm}dt\int^{\tau}_{0}ds \xi(t)L_{\text{Re}}(t-s)\xi(s) -\frac{1}{\hbar}\bar{S}_{\rm{eff}} \biggr]. \label{supp:chiWa}
\eeqa
Here, we define
\beq
\lambda_{1}(t)=\left\{ \begin{array}{ll}0 & \text{ if } t\leq \hbar\nu \\ v(t-\hbar\nu) & \text{ if }\hbar\nu\leq t \end{array} \right. ,\ \lambda_{2}(t)=\left\{ \begin{array}{ll} vt &\text{ if } t\leq \tau \\ v\tau & \text{ if }\tau\leq t \end{array} \right. .
\eeq
Now the integration over $DX$ will determine the functional form of $\xi(t)$ by solving the following differential equation:
\beq
 \ddot{\xi}(t)-\gamma\dot{\xi}(t)+\omega^{2}\xi(t)-\omega^{2}(\lambda_{1}(t)-\lambda_{2}(t))=0. \label{supp:xiequation}
\eeq
Here, we set the condition $\xi(\tau+\hbar\nu)=0$ which comes from $\delta(x_{f}-y_{f})$ inside the definition of the characteristic function of work. The solution to~(\ref{supp:xiequation}) is 
\beqa
\xi(t)&=&\frac{1}{\sin\omega_{\mathrm{d}}(\tau+\hbar\nu)}\biggl( \xi_{i}e^{\frac{\gamma}{2}t}\sin\omega_{\mathrm{d}}(\tau+\hbar\nu-t) \nonumber \\
& &-\omega^{2}e^{\frac{\gamma}{2}t}\int^{\tau+\hbar\nu}_{t}\hspace{-1mm}ds \Bigl(  \sin\omega_{\mathrm{d}}t \sin\omega_{\mathrm{d}}(\tau+\hbar\nu-s)+\sin\omega_{\mathrm{d}}(\tau+\hbar\nu-t)\sin\omega s  \Bigr)e^{-\frac{\gamma}{2}s}(\lambda_{1}(s)-\lambda_{2}(s))\biggr), \label{supp:xia}
\eeqa
with $\omega_{\mathrm{d}}=\sqrt{\omega^{2}-\gamma^{2}/4}$ (we consider the underdamped regime $\omega\geq \gamma/2$ in Fig. 2). Next, the integral over $dX_{f}$ in Eq.~(\ref{supp:chiWa}) will lead to $\dot{\xi}(\tau+\hbar\nu)=0$ (and also gives a constant which cancels $I_{0}$ in Eq.~(\ref{supp:chiWa})). We note that the condition $\dot{\xi}(\tau+\hbar\nu)=0$ determines $\xi_{i}$: 
\beq
\xi_{i}=\frac{\omega^{2}}{\omega_{\mathrm{d}}}\int^{\tau+\hbar\nu}_{0}\hspace{-1mm}ds e^{-\frac{\gamma}{2}s}(\lambda_{1}(s)-\lambda_{2}(s)) .\label{supp:xii}
\eeq
By substituting $\xi_{i}$ into Eq.~(\ref{supp:xia}), we obtain
\beq
\xi(t)=\frac{\omega^{2}}{\omega_{\mathrm{d}}}e^{\frac{\gamma}{2}t}\int^{\tau+\hbar\nu}_{t}\hspace{-1mm}ds e^{-\frac{\gamma}{2}s}(\lambda_{1}(s)-\lambda_{2}(s)) \sin\omega_{\mathrm{d}}(t-s). \label{supp:xi}
\eeq
We finally integrate over $dX_{i}$ in Eq.~(\ref{supp:chiWa}) and obtain 
\beqa
\chi_{W}(\nu)&=&\exp\biggl[-\frac{M\kappa}{2\beta\hbar^{2}}(\dot{\xi}(0)-\gamma\xi_{i})^{2}  -\frac{M\gamma}{2\hbar\pi}\int^{\omega_{D}}_{0}d\Omega \frac{\Omega\coth\frac{\hbar\Omega\beta}{2}}{(\omega^{2}-\Omega^{2})^{2}+\gamma^{2}\Omega^{2}}\biggl\{ \Omega^{2} \xi_{i}^{2} \nonumber \\
& & - 2\int^{\tau+\hbar\nu}_{0}\hspace{-1mm}dt\xi(t) \Bigl(  \xi_{i}(\Omega(\Omega^{2}-\omega^{2}-\gamma^{2})\sin\Omega t+\gamma(\omega^{2}-2\Omega^{2})\cos\Omega t \Bigr)-\dot{\xi}(0)\Bigl( (\omega^{2}-\Omega^{2})\cos\Omega t-\gamma\Omega\sin\Omega t\Bigr) \biggr\}  \nonumber \\
& &-\frac{1}{\hbar}\int^{\tau+\hbar\nu}_{0}\hspace{-1mm}dt\int^{t}_{0}ds \xi(t)L_{\text{Re}}(t-s)\xi(s) -\frac{iM\omega^{2}}{2\hbar}\int^{\tau+\hbar\nu}_{0}\hspace{-1mm}dt (\lambda_{1}^{2}(t)-\lambda_{2}^{2}(t))  \nonumber \\
&+&\frac{iM\omega^{4}}{2\hbar\omega_{\mathrm{d}}}\int^{\tau+\hbar\nu}_{0}\hspace{-1.5mm}dt \int^{\tau+\hbar\nu}_{t}\hspace{-1.5mm}ds e^{\frac{\gamma}{2}(t-s)}\sin\omega_{\mathrm{d}}(t-s)(\lambda_{1}(t)+\lambda_{2}(t))(\lambda_{1}(s)-\lambda_{2}(s)) \biggr].  \label{supp:solution}
\eeqa
Now the characteristic function of work depends only on $\xi(t)$, $\dot{\xi}(0)$ and $\xi_{i}$, which are uniquely determined by Eqs.~(\ref{supp:xii}) and (\ref{supp:xi}). Therefore, Eq.~(\ref{supp:solution}) gives the analytical expression for the characteristic function of work for a dragged harmonic oscillator.  


It is possible to simplify the imaginary part of Eq.~(\ref{supp:solution}) by explicitly calculating the integrals:
\beqa
& &\text{Im}[\ln\chi_{W}(\nu)] =M\gamma v^{2}(\nu\tau-\frac{\hbar\nu^{2}}{2})  \label{supp:chiim}\\
& &-\frac{M\gamma v^{2}}{\hbar\omega^{4}} \left(\omega^{2}-\frac{\gamma^{2}}{2}\right)\left( 2(e^{-\frac{\gamma}{2}\hbar\nu}\cos\omega_{\mathrm{d}}\hbar\nu-1)-e^{-\frac{\gamma}{2}(\tau+\hbar\nu)}\cos\omega_{\mathrm{d}}(\tau+\hbar\nu)+e^{-\frac{\gamma}{2}(\tau-\hbar\nu)}\cos\omega_{\mathrm{d}}(\tau-\hbar\nu)\right) \nonumber \\
& &+\frac{Mv^{2}}{2\hbar\omega^{4}\omega_{\mathrm{d}}}\left(\omega^{4}-2\omega^{2}\gamma^{2}+\frac{\gamma^{4}}{2}\right)\left(2e^{-\frac{\gamma}{2}\hbar\nu}\sin\omega_{\mathrm{d}}\hbar\nu-e^{-\frac{\gamma}{2}(\tau+\hbar\nu)}\sin\omega_{\mathrm{d}}(\tau+\hbar\nu)+e^{-\frac{\gamma}{2}(\tau-\hbar\nu)}\sin\omega_{\mathrm{d}}(\tau-\hbar\nu)\right) . \nonumber 
\eeqa
In the high-temperature limit, the real part of Eq.~(\ref{supp:solution}) can be further simplified by noting that $L_{\text{Re}}(t-s)=2M\gamma\beta^{-1}\delta(t-s)$, and Eq.~(\ref{supp:reducedrho}) reduces to the canonical distribution of the bare system. We then obtain
\beq
\text{Re}[\ln\chi_{W}(\nu)]= -\frac{M\gamma}{\hbar^{2}\beta}\int^{\tau+\hbar\nu}_{0}\hspace{-1mm}dt \xi^{2}(t) -\frac{M}{2\hbar^{2}\beta\omega^{2}}\Bigl(\omega^2\xi_{i}^{2}+ (\dot{\xi}(0)-\gamma\xi_{i})^2\Bigr)  \label{supp:chire}.
\eeq
We plot Fig.~3 in the main text by using Eqs.~(\ref{supp:chiim}) and (\ref{supp:chire}).

We finally note that the classical characteristic function of work is given by~\cite{Pan}
\beq
\chi_{W}(\nu)=\exp\biggl[\Bigl( i\nu-\frac{\nu^{2}}{\beta}\Bigr) \frac{Mv^{2}}{\omega^{2}}\Bigl(\gamma\tau\omega^{2}+(\gamma^{2}-\omega^{2})(e^{-\frac{\gamma\tau}{2}}\cos\omega_{\mathrm{d}}\tau-1)+\frac{\gamma}{2\omega_{\mathrm{d}}}(\gamma^{2}-3\omega^{2})e^{-\frac{\gamma\tau}{2}}\sin\omega_{\mathrm{d}}\tau\Bigr) \biggr]. \label{supp:classical}
\eeq
It can be checked that when $\hbar\rightarrow 0$ and $\beta\rightarrow 0$, Eq.~(\ref{supp:solution}) converges to Eq.~(\ref{supp:classical}), which is a demonstration of the quantum-classical correspondence of the characteristic function of work in the dragged harmonic oscillator. We would like to emphasize that the work distribution from Eq.~(\ref{supp:solution}) is non Gaussian but the work distribution from Eq.~(\ref{supp:classical}) is Gaussian. This result is similar to the dragged harmonic oscillator in the isolated regime~\cite{Talkner08}.


\begin{thebibliography}{99}


\bibitem{FeynmanHibbs} R. P. Feynman and A. R. Hibbs, {\it Quantum Mechanics and Path integrals,} edited by D. F. Styer, (Dover, New York, 2010).

\bibitem{Rajaraman} R. Rajaraman, {\it Solitons and instantons,} (Amsterdam: North Holland, 1987).


\bibitem{Fujikawa} K. Fujikawa, {\it Path-Integral Measure for Gauge-Invariant Fermion Theories,} Phys. Rev. Lett. {\bf 42,} 1195 (1979).

\bibitem{Sondhi} S. L. Sondhi, S. M. Girvin, J. P. Carini, and D. Shahar, {\it Continuous quantum phase transitions,} Rev. Mod. Phys. {\bf 69,} 315 (1997).


\bibitem{Feynman} R. P. Feynman and F. L. Vernon, Jr., {\it The Theory of a General Quantum System Interacting with a Linear Dissipative System,} Ann. Phys. (N. Y.) {\bf 24,} 118 (1963).

\bibitem{Caldeira83} A. O. Caldeira and A. J. Leggett, {\it  Path Integral Approach to Quantum Brownian Motion,} Physica A {\bf 121,} 587 (1983).

\bibitem{Hanggiaspects} P. Talkner and P. Hanggi, {\it Aspects of quantum work}, Phys. Rev. E {\bf 93,} 022131 (2016).

\bibitem{Esposito} M. Esposito, U. Harbola, S. Mukamel, {\it Nonequilibrium fluctuations, fluctuation theorems, and counting statistics in quantum systems}, Rev. Mod. Phys. {\bf 81}, 1665 (2009).

\bibitem{fluctuation1} M. Campisi, P. H\"{a}nggi and P. Talkner, {\it Colloquium: Quantum fluctuation relations: Foundations and applications}, Rev. Mod. Phys. {\bf 83,} 771 (2011),  {\it erratum:} {\bf 83,} 1653 (2011)


\bibitem{Pekola} J. P. Pekola, {\it Towards quantum thermodynamics in electronic circuits.} {\it Nat. Phys.} {\bf 11}, 118 (2015).

\bibitem{Anders} S. Vinjanampathy and J. Anders, {\it Quantum Thermodynamics,} Contemporary Physics, {\bf 57,} 545 (2016).

\bibitem{Strasberg} P. Strasberg, G. Schaller, T. Brandes, and M. Esposito, {\it Quantum and Information Thermodynamics: A Unifying Framework Based on Repeated Interactions.} Phys. Rev. X {\bf 7,} 021003 (2017).

\bibitem{CaldeiraBook} A. O. Caldeira, {\it An Introduction to Macroscopic Quantum Phenomena and Quantum Dissipation,} (Cambridge University Press, Cambridge, UK, 2014).

\bibitem{Sun94} L. H. Yu and C. P. Sun, {\it Evolution of the Wave Function in a Dissipative System,} Phys. Rev. A {\bf 49,} 592 (1994).

\bibitem{Hanggi05C} P. H\"{a}nggi and G.-L. Ingold, {\it Fundamental Aspects of Quantum Brownian Motion}, Chaos {\bf 15,} 026105 (2005)


\bibitem{Ueda} M. Ueda, {\it Transmission spectrum of a tunneling particle interacting with dynamical fields: Real-time functional-integral approach}, Phys. Rev. B {\bf 54,} 8676 (1996).

\bibitem{Saito} K. Saito and A. Dhar, {\it Fluctuation Theorem in Quantum Heat Conduction,} Phys. Rev. Lett. {\bf 99,} 180601 (2007).

\bibitem{Kato} A. Kato and Y. Tanimura, {\it Quantum heat current under non-perturbative and non-Markovian conditions: Applications to heat machines,} J. Chem. Phys. {\bf 145,} 224105 (2016).


\bibitem{coherence} M. O. Scully, M. S. Zubairy, G. S. Agarwal, H. Walther, {\it Extracting Work from a Single Heat Bath via Vanishing Quantum Coherence,} Science {\bf 299,} 862-864 (2003).

\bibitem{Dong} Y. Dong, K. Zhang, F. Bariani, and P. Meystre, {\it Work measurement in an optomechanical quantum heat engine,} Phys. Rev. A {\bf 92,} 033854 (2015).



\bibitem{Pekola16} B. Karimi and J. P. Pekola, {\it Otto refrigerator based on a superconducting qubit: Classical and quantum performance,} Phys. Rev. B {\bf 94,} 184503 (2016).


\bibitem{quantumjarexp} S. An, J.-N. Zhang, M. Um, D. Lv, Y. Lu, J. Zhang, Z.-Q. Yin, H. T. Quan and K. Kim, {\it Experimental test of the quantum Jarzynski equality with a trapped-ion system}, Nature Phys. {\bf 11}, 193 (2015).


\bibitem{Benjamin} N. Cottet, S. Jezouin, L. Bretheau, P. C.-Ibarcq, Q. Ficheux, J. Anders, A. Auff\`{e}ves, R. Azouit, P. Rouchon, and B. Huard, {\it Observing a quantum Maxwell demon at work,} Proc. Natl. Acad. Sci. {\bf 114,} 7561 (2017).

\bibitem{Masuyama} Y. Masuyama, K. Funo, Y. Murashita, A. Noguchi, S. Kono, Y. Tabuchi, R. Yamazaki, M. Ueda, and Y. Nakamura, {\it Information-to-work conversion by Maxwell's demon in a superconducting circuit-QED system}, Nat. Commun. {\bf 9,} 1291 (2018).

\bibitem{Parrondo} J. M. R. Parrondo, J. M. Horowitz and T. Sagawa, {\it Thermodynamics of information}, Nat. Phys. {\bf 11}, 131 (2015).

\bibitem{Horodecki} M. Horodecki and J. Oppenheim, {\it Fundamental limitations for quantum and nanoscale thermodynamics},  Nat. Commun. {\bf 4}, 2059 (2013).

\bibitem{Campisi09} M. Campisi, P. Talkner, and P. H\"{a}nggi, {\it Fluctuation Theorem for Arbitrary Open Quantum Systems,} Phys. Rev. Lett. {\bf 102,} 210401 (2009).


\bibitem{Horowitz1} J. M. Horowitz, {\it Quantum-trajectory approach to the stochastic thermodynamics of a forced harmonic oscillator}, Phys. Rev. E {\bf 85}, 031110 (2012).


\bibitem{Hekking} F. W. J. Hekking and J. P. Pekola, {\it Quantum Jump Approach for Work and Dissipation in a Two-Level System}, Phys. Rev. Lett. {\bf 111}, 093602 (2013).

\bibitem{Liu1} F. Liu, {\it Calculating work in adiabatic two-level quantum Markovian master equations: A characteristic function method}, Phys. Rev. E {\bf 90,} 032121 (2014).

\bibitem{Suomela} S. Suomela, A. Kutvonen, and T. Ala-Nissila, {\it Quantum jump model for a system with a finite-size environment,} Phys. Rev. E {\bf 93,} 062106 (2016). 







\bibitem{Sekimoto} K. Sekimoto, {\it Stochastic Energetics} (Lecture Notes in Physics vol 799), Springer-Verlag Berlin Heidelberg, (2010).

\bibitem{Sekimoto98} K. Sekimoto, {\it Langevin Equation and Thermodynamics.} Prog. Theo. Phys. Supp. {\bf 130,} 17 (1998).

\bibitem{Seifert} U. Seifert, {\it Stochastic thermodynamics,  fluctuation theorems and molecular machines.} Rep. Prog. Phys. {\bf 75}, 126001 (2012).

\bibitem{Stochasticbook}  R. Klages, W. Just, and C. Jarzynski, eds., {\it Nonequilibrium Statistical Physics of Small Systems: Fluctuation Relations and Beyond} (Wiley-VCH, 2013).

\bibitem{Jarzynski1} C. Jarzynski, {\it Nonequilibrium equality for free energy differences.} Phys. Rev. Lett. {\bf 78}, 2690 (1997).

\bibitem{Jarzynski2} C. Jarzynski, {\it Equilibrium free-energy differences from nonequilibrium measurements: A master-equation approach.} Phys. Rev. E.{\bf 56}, 5018 (1997).

\bibitem{Crooks} G. E. Crooks, {\it Entropy production fluctuation theorem and the nonequilibrium work relation for free energy differences.}  Phys. Rev. E {\bf 60}, 2721-2726 (1999).

\bibitem{Crooks00} G. E. Crooks, {\it Path-ensemble averages in systems driven far from equilibrium.} Phys. Rev. E {\bf 61,} 2361 (2000).

\bibitem{Hummer} G. Hummer and A. Szabo, {\it Free energy reconstruction from nonequilibrium single-molecule pulling experiments.} Proc. Natl. Acad. Sci. U.S.A. {\bf 98,} 3658 (2001).

\bibitem{Onsarger} L. Onsager and S. Machlup, {\it Fluctuations and Irreversible Processes.} Phys. Rev. {\bf 91,} 1505 (1953). 

\bibitem{cpath0} C. Jarzynski, {\it Equilibrium Free Energies from Nonequilibrium Processes.} Acta Physica Polonica B {\bf 29,} 1609 (1998).

\bibitem{cpath1} V. Y. Chernyak, M. Chertkov and C. Jarzynski, {\it  Path-integral analysis of fluctuation theorems for general Langevin processes.} J. Stat. Mech. P08001 (2006).

\bibitem{cpath2} T. Taniguchi and E. G. D. Cohen, {\it Inertial effects in nonequilibrium work fluctuations by a path integral approach.}
J. Stat. Phys. {\bf 130,} 1 (2008).

\bibitem{cpath3} D. D. L. Minh and A. B. Adib, {\it Path integral analysis of Jarzynski's equality: Analytical results.} Phys. Rev. E  {\bf 79,} 021122 (2009).


\bibitem{Kac} M. Kac, {\it Wiener and integration in function spaces.} Bull. Am. Math. Soc., {\bf 72,} 52, (1966).

\bibitem{Kurchan} J. Kurchan, {\it A Quantum Fluctuation Theorem.} arXiv:cond-mat/0007360.

\bibitem{Tasaki} H. Tasaki, {\it Jarzynski Relations for Quantum Systems and Some Applications.} arXiv:cond-mat/0009244.


\bibitem{Weiss} U. Weiss, {\it Quantum Dissipative Systems.} (World Scientific, Singapore, 2012).

\bibitem{Caldeira87} C. Morais Smith and A. O. Caldeira, {\it Generalized Feynman-Vernon approach to dissipative quantum systems.} Phys. Rev. A {\bf 36,} 3509 (1987).

\bibitem{Grabert} H. Grabert, P. Schramm, and G.-L. Ingold, {\it Quantum Brownian Motion: The Functional Integral Approach.} Phys. Rep. {\bf 168,} 115 (1988).

\bibitem{Tanimura} Y. Tanimura, {\it Stochastic Liouville, Langevin, Fokker-Planck, and Master Equation Approaches to Quantum Dissipative Systems.} J. Phys. Soc. Jpn. {\bf 75,} 082001 (2006).






\bibitem{Jarzynski04} C. Jarzynski, {\it Nonequilibrium work theorem for a system strongly coupled to a thermal environment.} J. Stat. Mech. P09005 (2004).

\bibitem{Seifert16} U. Seifert, {\it First and Second Law of Thermodynamics at Strong Coupling.} Phys. Rev. Lett. {\bf 116,} 020601 (2016).

\bibitem{Jarzynski17} C. Jarzynski, {\it Stochastic and Macroscopic Thermodynamics of Strongly Coupled Systems.} Phys. Rev. X {\bf 7,} 011008 (2017).

\bibitem{Talkner16} P. Talkner and P. H\"{a}nggi, {\it Open system trajectories specify fluctuating work but not heat.} Phys. Rev. E. {\bf 94,} 022143 (2016).









\bibitem{Hu} Y. Subasi and B. L. Hu, {\it Quantum and classical fluctuation theorems from a decoherent histories, open-system analysis.} Phys. Rev. E {\bf 85}, 011112 (2012).


\bibitem{Carrega15} M. Carrega, P. Solinas, A. Braggio, M. Sassetti and U. Weiss, {\it Functional integral approach to time-dependent heat exchange in open quantum systems: general method and applications.} New. J. Phys. {\bf 17,} 045030 (2015).

\bibitem{Carrega16} M. Carrega, P. Solinas, M. Sassetti and U. Weiss, {\it Energy Exchange in Driven Open Quantum Systems at Strong Coupling.} Phys. Rev. Lett. {\bf 116,} 240403 (2016).

\bibitem{Aurell} E. Aurell and R. Eichhorn, {\it On the von Neumann entropy of a bath linearly coupled to a driven quantum system.} New. J. Phys. {\bf 17,} 065007 (2015).

\bibitem{Aurell17} E. Aurell, {\it On Work and Heat in Time-Dependent Strong Coupling.} {\it Entropy} {\bf 19,} 595 (2017).

\bibitem{characteristic1} P. Talkner, E. Lutz and P. H\"{a}nggi, {\it Fluctuation theorems: Work is not an observable.} Phys. Rev. E {\bf 75} 050102 (2007).



\bibitem{Ramsey} R. Dorner, S. R. Clark, L. Heaney, R. Fazio, J. Goold, and V. Vedral, {\it Extracting Quantum Work Statistics and Fluctuation Theorems by Single-Qubit Interferometry.} Phys. Rev. Lett. {\bf 110,} 230601 (2013).

\bibitem{identity} The identity $(i/\hbar)S_{1}^{\nu}[x]=(i/\hbar)S_{2}^{\nu}[x]+i\nu W_{\nu}[x]$ can be easily shown by noting that $\partial_{\tau}S^{\nu}_{1}[x]=\partial_{\tau}S^{\nu}_{2}[x]+\hbar\nu\partial_{\tau}W_{\nu}[x]$ and $S^{\nu}_{1}[x]|_{\tau=0}=S^{\nu}_{2}[x]|_{\tau=0}+\hbar\nu W_{\nu}[x]|_{\tau=0}$.







\bibitem{commenta} The work functional~(\ref{WFunction}) vanishes when $\dot{\lambda}_{t}=0$ because the work should only depend on the energy supplied from the time-dependent variation of the external control. Apart from this constraint, there is a freedom of choosing the form of the phase and the work functional in Eq.~(\ref{isoWCFtwo}).  However, when calculating the quantity $\langle W^{n}\rangle$ in Eq.~(\ref{Wfinitemoment}), the choice does not influence the result.


\bibitem{commentaa} The $\nu$ derivative acting on $e^{\frac{i}{\hbar}(\Delta S^{\nu}_{2}[x]-\Delta S^{\nu}_{2}[y])}$ with $\Delta S^{\nu}_{2}[y]:=\int^{\tau+\hbar\nu}_{\tau}ds \mathcal{L}[\lambda_{\tau},y(s)]$ will vanish because of the delta function $\delta(x_{f}-y_{f})$ in $\chi_{W}(\nu)$. Also note that $\Delta S^{\nu}_{2}|_{\nu=0}=0$. Therefore, $\Delta S^{\nu}_{2}$ does not appear in the formula for $\langle W^{n}\rangle$. 



\bibitem{commentclassicallimit} See Sec.~I of Ref.~\cite{supplement} for details of the classical limit of the characteristic function of work.

\bibitem{supplement} Supplementary material. The supplementary material includes Ref.~\cite{Pan}.

\bibitem{Pan} R. Pan, T. M. Hoang, Z. Fei, T. Qiu, J. Ahn, T. Li, and H. T. Quan, {\it The validity and breakdown of the overdamped approximation in stochastic thermodynamics: Theory and experiment}, arXiv:1805.09080.

\bibitem{QCcorrespondence1} C. Jarzynski, H. T. Quan, and S. Rahav, {\it Quantum-Classical Correspondence Principle for Work Distributions,} Phys. Rev. X {\bf 5,} 031038 (2015).

\bibitem{QCcorrespondence2} L. Zhu, Z. Gong, B. Wu, and H. T. Quan, {\it Quantum-classical correspondence principle for work distributions in a chaotic system,} Phys. Rev. E {\bf 93,} 062108 (2016).




\bibitem{Caldeira83b} A. O. Caldeira and A. J. Leggett, {\it Quantum Tunnelling in a Dissipative System,} Ann. Phys. {\bf 149,} 374 (1983).



\bibitem{footnoteJar} See Sec. II of Ref.~\cite{supplement} for the derivation of Jarzynski's equality using the path integral expression.

\bibitem{Talkner08} P. Talkner, P. S. Burada and P. Hanggi, {\it Statistics of work performed on a forced quantum oscillator,} Phys. Rev. E {\bf 78,} 011115 (2008).

\bibitem{Caldeira90} C. M. Smith and A. O. Caldeira, {\it Application of the generalized Feynman-Vernon approach to a simple system: The damped harmonic oscillator,} Phys. Rev. A {\bf 41,} 3103 (1990).





\bibitem{Supplement} See Sec. III of Ref.~\cite{supplement} for the details of the analytical calculation for a dragged harmonic oscillator. 



























\end{thebibliography}

\begin{thebibliography}{99}

\bibitem{characteristic1} P. Talkner, E. Lutz and P. H\"{a}nggi, {\it Fluctuation theorems: Work is not an observable}, Phys. Rev. E {\bf 75} 050102 (2007).

\bibitem{Campisi09} M. Campisi, P. Talkner, and P. H\"{a}nggi, {\it Fluctuation Theorem for Arbitrary Open Quantum Systems,} Phys. Rev. Lett. {\bf 102,} 210401 (2009).


\bibitem{Talkner08} P. Talkner, P. S. Burada and P. Hanggi, {\it Statistics of work performed on a forced quantum oscillator}, Phys. Rev. E {\bf 78,} 011115 (2008).

\bibitem{Caldeira90} C. M. Smith and A. O. Caldeira, {\it Application of the generalized Feynman-Vernon approach to a simple system: The damped harmonic oscillator}, Phys. Rev. A {\bf 41,} 3103 (1990).

\bibitem{Pan} R. Pan, T. M. Hoang, Z. Fei, T. Qiu, J. Ahn, T. Li, and H. T. Quan, {\it The validity and breakdown of the overdamped approximation in stochastic thermodynamics: Theory and experiment}, arXiv:1805.09080.

\end{thebibliography}
\end{document}